\documentclass[aps,prd,showpacs,twocolumn,nofootinbib,superscriptaddress]{revtex4-2}
\usepackage{amsmath,amssymb,amsfonts}
\usepackage{bm}
\usepackage{verbatim}
\usepackage{hyperref}
\usepackage{slashed,braket}
\usepackage{ulem}
\usepackage{cancel}
\usepackage{color}
\usepackage{graphicx,graphics}
\usepackage[title,titletoc]{appendix}
\usepackage{stackengine}
\usepackage{mathrsfs}
\usepackage{tikz-feynman}
\usepackage{amsthm}

\usepackage{tikz}
\usepackage{pgfplots}

\usepackage{lineno}
\usepackage[T1]{fontenc} 
\usepackage{braket}

\usepackage{bbold}
\usepackage{comment}
\excludecomment{confidential}
\usepackage{makecell}
\usepackage{bm}
\usepackage{braket}
\usepackage{slashed}
\usepackage{mathtools}

\usepackage{amsthm}

\newcommand{\mz}[1]{{#1}}

\newcommand{\Tr}{\operatorname{Tr}}

\newcommand{\tr}{\operatorname{tr}}

\begin{document}
	
	\title{Topological invariant responsible for the stability of the Fermi surfaces in non - homogeneous systems}

	\author{M.A. Zubkov}
	\email{mikhailzu@ariel.ac.il}
	\affiliation{Physics Department, Ariel University, Ariel 40700, Israel}

	\date{\today}
	
	\begin{abstract}
		The topological invariant responsible for the stability of Fermi point/Fermi surface in homogeneous systems is expressed through the one particle Green function, which depends on momentum. It is given by an integral over the 3D hypersurface in momentum space surrounding the Fermi surface. Notion of Fermi surface may be extended to the non - homogeneous systems using Wigner - Weyl calculus. The Fermi surface becomes coordinate dependent, it may be defined as the position of the singularity in momentum space of the Wigner transformed Green function. Then the topological invariant responsible for the stability of this Fermi surface is given by the same expression as for the homogeneous case, in which the Green function is replaced by its Wigner transformation while the ordinary products are replaced by the Moyal products. We illustrate the proposed construction by the examples of the systems, in which the given topological invariant is nontrivial and may be calculated explicitly.
	\end{abstract}
	
	\maketitle

\section{Introduction}
The topological invariant responsible for the quantum Hall effect is widely known as the  TKNN invariant \cite{Thouless1982}. It takes integer values  \cite{Kaufmann:2015lga}. For the discussion of the topology associated with it see also \cite{Avron1983,Fradkin1991,Tong:2016kpv,Hatsugai1997,Qi2008}. In the presence of interactions it is not defined, and for the intrinsic anomalous QHE it is to be replaced by expression through the two - point Green functions  
\begin{equation}
	{\cal N}
=  \frac{ \epsilon_{ijk}}{  \,3!\,4\pi^2}\, \int d^3p \Tr
\Bigl[
{G}(p ) \frac{\partial {G}^{-1}(p )}{\partial p_i}  \frac{\partial  {G}(p )}{\partial p_j}  \frac{\partial  {G}^{-1}(p )}{\partial p_k}
\Bigr].
\label{cal0}
\end{equation}
See \cite{IshikawaMatsuyama1986,Volovik1988,Volovik2003a,parity_anomaly,parity_anomaly_} and  \cite{Zubkov2018a,ZZ2019}. Extention to the multi - dimensional space - time was discussed in  \cite{mera2017topological}. Even more simple topological invariant composed of the two-point Green  function is responsible for the stability of the Fermi surface in the homogeneous $3+1$ D systems:
\begin{equation}
N_1= \tr \oint_C \frac{1}{2\pi i}
G(p_0,\bar{p})d G^{-1}(p_0,\bar{p}).
\end{equation}
Here $C$ is a contour, which windes ones around the Fermi surface \cite{Volovik2003a} in four-dimensional momentum space.
The topological stability of Fermi points is protected by expression similar to that of Eq. (\ref{cal0})  \cite{Matsuyama1987a,Volovik2003a}
\begin{eqnarray}
N_3=\frac1{24\pi^2} \epsilon_{\mu\nu\rho\lambda} \tr\int_S dS^\mu
G\partial^\nu G^{-1}
G\partial^\rho G^{-1}
G\partial^\lambda G^{-1}.\label{cal1}
\end{eqnarray}
Here $S$ is the surface that surrounds the given Fermi point. If chemical potential deviates from the position of the Fermi point, the latter is transformed to a Fermi surface. In this situation Eq. (\ref{cal1}) protects its stability as a whole: it can change its form, and even shrink to a point, but it cannot disappear completely without a phase transition. 

If in addition to the interations the inhomogeneity is present in the system (in particular, due to the presence of disorder or external magnetic field), both Eq. (\ref{cal0}) and Eq. (\ref{cal1}) are undefined. In \cite{ZW2019,ZZ2022} it was shown that in such systems the topological invariant responsible for the QHE has the same form as Eq. (\ref{cal0}), but with averaging over the system sample, and with the two - point Green function replaced by its Wigner transform. The ordinary products are then replaced by the Moyal products:
\begin{eqnarray}
	{\cal N}
	&=&  \frac{ \epsilon_{ijk}}{  \,3!\,4\pi^2 |A|}\,\int d^2x\, \int d^3p \Tr
	\Bigl[
	{G}_W(x,p)\star \frac{\partial {Q}_W(x,p)}{\partial p_i} \nonumber\\&&\star \frac{\partial  {G}_W(x,p)}{\partial p_j} \star \frac{\partial  {Q}_W(x,p)}{\partial p_k}
	\Bigr].
	\label{cal01}
\end{eqnarray}
Here $\star$ is Moyal product while $|A|$ is the sample area. $G_W$ is the Wigner transformed complete interacting Green function, $Q_W$ is Weyl symbol of operator inverse to the two - point Green function.

In \cite{Khaidukov2017,Zubkov2017} it was shown that expression similar to that of Eq. (\ref{cal1}) is responsible for the chiral separation effect (CSE) in $3+1$D  relativistic systems and in the condensed matter systems with emergent relativistic invariance (i.e. the conductivity of the latter is proportional to it):
\begin{eqnarray}
	N_3&=&\frac1{48\pi^2} \epsilon_{\mu\nu\rho\lambda} \tr\,\gamma^5 \,\int_S dS^\mu
	G\partial^\nu G^{-1}\nonumber\\&&
	G\partial^\rho G^{-1}
	G\partial^\lambda G^{-1}.\label{cal11}
\end{eqnarray}

The further consideration of the CSE in the non - homogeneous systems 
\cite{SuleymanovZubkov2020,ZA2023,Z2024} has led to the conclusion that Eq. (\ref{cal11}) is to be extended further:
\begin{eqnarray}
	N_3&=&\frac1{48\pi^2|V|} \epsilon_{\mu\nu\rho\lambda} \tr\,\gamma^5 \,\int \,d^3x \,\int_S dS^\mu
	G_W \star \partial^\nu Q_W \star
	G_W \nonumber\\&&\star \partial^\rho  Q_W \star
	G_W\star \partial^\lambda  Q_W.\label{cal12}
\end{eqnarray}
Here $|V|$ is the three - volume of the system. It is assumed that $\gamma^5$ commutes or anticommutes with $Q_W$ and $G_W$ along the Fermi surface. It is important to notice that in Eq. (\ref{cal12}) surface $S$ must have the form of the two hyperplanes $p_4 = \pm \epsilon \to 0$ \mz{with opposite orientations} situated just below the position of Fermi energy and just above it. It was noticed in \cite{SuleymanovZubkov2020,ZA2023,Z2024} that these two hyperplanes may be safely deformed to the hypersurfaces of the other form surrounding the singularities of expression standing inside the integral over $p$. {\it The latter singularities actually represent the position of the Fermi surface depending on $x$. This is the way to extend the notion of the Fermi surface to the non - homogeneous systems.} However, the mentioned above deformation of hypersurface $S$ should be accompanied by a modification of expressions for $G_W$ and $\star$ (which was not mentioned in \cite{SuleymanovZubkov2020,ZA2023,Z2024}). \mz{Besides, surface $\Sigma_3(x)$ still should consist of the two infinite pieces. Here vector $\bar{p}$ of spatial momentum parametrizes this surface. } We already discussed this modification in \cite{XZ2024}. In the present paper we touch this issue in more details.  

In the present paper we extend the discussion of \cite{SuleymanovZubkov2020,ZA2023,Z2024} to several directions. First of all, as it was mentioned above, we demonstrate how the topological invariant of Eq. (\ref{cal12}) is to be defined correctly for the case, when hypersurface $S$ has \mz{a more general} form. Next, we consider the more general case of the topological invariants from the same family, which are responsible for the stability of the Fermi points/Fermi surfaces that are not protected by any symmetry, or which are protected by a symmetry different (in general case) from the chiral symmetry of Eq. (\ref{cal12}).

\section{Covariant Wigner-Weyl calculus}
In this section we briefly review the covariant Wigner - Weyl calculus for quantum field theory proposed recently in \cite{XZ2024}. It generalizes the previously used construction (see, for example, \cite{SuleymanovZubkov2020} and references therein).  For simplicity unlike in \cite{XZ2024} we consider external electromagnetic field rather than external non - Abelian gauge field. Below is a summary of basic results:
\begin{itemize}
	\item Let us work in 4 Euclidean spacetime dimensions.  Let $A_\mu(x)$ be an external $U(1)$ gauge field. Let us introduce a Hilbert space, $H$ over $R^4$. We use the ``bra-ket'' notation \cite{griffiths2018introduction} for states in $H$. The ``position and momentum operators'' will be denoted by $\hat x_\mu$ and $\hat p_\mu$  (i.e. $\hat x_\mu$ corresponds to $\psi(x)\to x_\mu\psi(x)$ and $\hat p_\mu$ corresponds to $\psi(x)\to-i\partial_\mu \psi(x)$). We also denote $\hat{\pi} = \hat{p} - A(\hat{x})$. We suppose that  there is internal space of dimension $M$. Let $\hat X$ be a $M\times M$ matrix of operators in the Hilbert space. The covariant Weyl symbol of $\hat X$ is defined as
	\begin{eqnarray}
		&&X_{\cal{W}}(x,p):=\int d^4y\,e^{ipy}\,\label{12fe}U(x,x-y/2)\\&& \bra{x-y/2}\hat{X}\ket{x+y/2} U(x+y/2,x)  \nonumber
	\end{eqnarray}
with the parallel transporter $U(w,z) = e^{i\int_z^w A(y)dy}$ along the straight line connecting the two points. 
\eqref{12fe} is to be compared with the ordinary Weyl symbol \cite{SuleymanovZubkov2020}
\begin{align}
	X_{{W}}(x,p):=&\int d^4y\,e^{ipy}\,\bra{x-y/2}\hat{X}\ket{x+y/2}   \label{12feo}
\end{align}	
	
	\item \eqref{12fe} can also be written as
	\begin{align}
		X_{\cal{W}}(x,p)&=\int d^4y\,e^{ipy}\,\bra{x}e^{-\frac{iy}{2}\hat{\pi}}\hat{X}e^{-\frac{iy}{2}\hat{\pi}}\ket{x}  \label{pi}
	\end{align}
while \eqref{12feo} is equivalent to
\begin{align}
	X_{{W}}(x,p)&=\int d^4y\,e^{ipy}\,\bra{x}e^{-\frac{i}{2}y\hat{p}}\hat X \,e^{-\frac{i}{2}y\hat{p}}\ket{x}  \label{pi}
\end{align}
	This follows from 
	\begin{align}
		e^{-iy\hat{ {\pi}}}\ket{x}&=\ket{x+y}U(x+y,x) \label{usin1}\\
		\text{and}\quad\bra{x}	e^{iy\hat{{\pi}}}&=U(x,x+y)\bra{x+y} \label{usin5}
	\end{align}
	and
	\begin{align}
		\exp(y D)\psi(x)
		&=U(x,x+y)\psi(x+y) \label{adjoint}
	\end{align}
where $D = \partial - i A$ is the covariant derivative.

	\item The inverse (Weyl) transform of \eqref{12fe} is
	\begin{eqnarray}
		\hat X&=&(2\pi)^{-8}\int d^4q d^4y d^4x d^4p \nonumber\\&& \,e^{\frac{i}{2}y(\hat{ {\pi}}-p)}e^{iq(\hat x-x)} {X}_{\cal{W}}(x,p)\,e^{\frac{i}{2}y(\hat{ {\pi}}-p)} \label{2}
	\end{eqnarray}
	
	
	\item For two operators $\hat X$ and $\hat Y$, the Moyal (star) product gives $X_W\star Y_W:=(\hat X\hat Y)_W$. The direct formula for it is 
	\begin{align}
		\begin{split}
			&(X_W\star Y_W)(x,p)\\
			=&(2\pi)^{-8}\int d^4yd^4k d^4y'd^4k'\,e^{-iy(k-p)-iy'(k'-p)}\times \\
			& X_W(x-y'/2,k) \,  \, Y_W(x+y/2,k') \, 
			\\
			=& X_W(x,p) \, e^{\frac{i}{2}(\overleftarrow{\partial}_{x}\overrightarrow{\partial}_{p}-\overleftarrow{\partial}_{p}\overrightarrow{\partial}_{x})} \, Y_W(x,p) \,
		\end{split} \label{starero}
	\end{align}
	For the details see, for example, 
 \cite{hillery1984distribution, moyal1949quantum,zachos2005quantum}.
	 In the same way the covariant Moyal product may be defined as $X_{\cal W}\bigstar Y_{\cal W}:=(\hat X\hat Y)_{\cal W}$, which gives 
	 \begin{align}
	 	\begin{split}
	 		&(X_{\cal W}\bigstar Y_{\cal W})(x,p)\\
	 		&=(2\pi)^{-8}\int d^4yd^4k d^4y'd^4k'\,\\&e^{-iy(k-p)-iy'(k'-p)}\times \\
	 		& X_{\cal W}(x-y'/2,k) \,  \, Y_{\cal W}(x+y/2,k') \,\\
	 		&  U(x+(y+y')/2, x-(y+y')/2) \\& U(x-(y+y')/2 , x-y'/2+y/2)  \\ &   U(x-y'/2+y/2 ,x+(y+y')/2) 
	 		\end{split} 
	 		 \label{starer}
	 \end{align}
	 
	\item Denoting trace over the Hilbert space by $\text{tr}_H$ we have
	\begin{align}
	 {\rm tr}_H (\hat{X} \hat{Y})&= (2\pi)^{-4}\int d^4x d^4p \,{\rm tr}_G (X_W  Y_W) \\&=(2\pi)^{-4}\int d^4x d^4p \,{\rm tr}_G (X_{\cal W}  Y_{\cal W})\label{TrXY}
	\end{align}
	{Also, in the following, we will denote by $\tr_D$ the trace over spinor indices and over the $M$-dimensional extra
		internal space.}
	
	
	\item 
	{(\ref{starer}) allows us to derive the following identities:
		\begin{eqnarray}
			&&\int \frac{d^4x d^4p}{(2\pi^4)}	{\rm tr_G}(X_{\cal W}(x,p)\bigstar Y_{\cal W}(x,p)) =\nonumber\\&&= \int \frac{d^4x d^4p}{(2\pi^4)}	{\rm tr}_G(X_{\cal W}(x,p) Y_{\cal W}(x,p)) = \nonumber\\&&=\int \frac{d^4x d^4p}{(2\pi^4)}	{\rm tr}_G(Y_{\cal W}(x,p)\bigstar X_{\cal W}(x,p)) \label{cyclTr}
		\end{eqnarray}
		provided that these integrals are convergent. The similar expression is valid for the ordinary Moyal product with the covariant star product $\bigstar$ substituted by the ordinary Moyal product $\star$.}
	
\end{itemize}

\section{Quantum field theory in terms of Wigner-Weyl calculus}

Let $\psi(x)$ denote a two-component Weyl spinor. The partition function of the model is
\begin{align}
	Z=&\int D\bar\psi D\psi\,e^S \label{pf} \\
	\text{with}\quad S=&\int d^4x \,\bar\psi(x)Q(x,-iD)\psi(x)\label{ac}\\
	\text{and}\quad Q(x,-iD)=&\sum_{|\alpha|\leq m}  c_\alpha(x) (-iD)^\alpha \label{sym}
\end{align}
\eqref{sym} is written in the multi-index notation
with the multi-index $\alpha=(\alpha_1,\alpha_2,\alpha_3,\alpha_4)$, $|\alpha| := \sum_\mu \alpha_\mu$ and 
$(-i D)^\alpha := \prod_\mu (-i D_\mu)^{\alpha_\mu}$. $m$ is the {order} of $Q$ and it is a positive integer. $D_\mu:=\partial_\mu-iA_\mu$ is the covariant derivative. $c_{\alpha}(x)$ is a {$M\times M$} matrix of functions.
We require that  $Q$ is elliptic.
The ``inhomogeneity'' in this model comes from the explicit $x$ dependence in $Q(x,-iD)$. However, dependence on $x_4$ does not present because we deal with the equilibrium theory. 

In the Hilbert space operator notation \eqref{ac} can be written as 
\begin{align}
	S&=-\text{tr}_{D}\text{tr}_{H}\left(\hat Q\hat\rho\right)\\
	\text{where}\quad\hat Q&:=Q(\hat x,\hat \pi), \quad \hat{\pi} = \hat{p} - A(\hat{x})\\
	\text{and}\quad\bra{x}\hat{\rho}\ket{y}&:=\psi(x)\bar\psi(y) \label{usin}
\end{align}
where $\text{tr}_D$,  and $\text{tr}_H$ are traces w.r.t. the $M$ internal space indices, and the
	Hilbert space respectively.
	
Let us calculate the Weyl symbol of operator $\hat{Q}$. For this purpose let us represent operator $Q(x,-iD)$ in the following form 
	\begin{equation}
		Q(x,-iD) = \sum_{|\alpha| \le m} o_{\alpha}(x) \circ (-i D)^\alpha \label{sym0}
	\end{equation}
Here coefficients $o_\alpha$ differ from the introduced above coefficients $c_\alpha$, and may be expressed through the latter (as well as through the derivatives of $c_\alpha$) due to the commutation relations between $\hat{\pi}$ and $c_{\alpha}(\hat{x})$. 
	 By $o_{\alpha}(x) \circ (-i D)^\alpha$ we denote 
	\begin{eqnarray}
		&&o_{\alpha}(x) \circ (-i D)^\alpha = \frac{1}{2^{|\alpha|}} \{...\{o_{\alpha}(x),(-i D_1)\}...\nonumber\\&&...(-i D_1)\}(-i D_2)\}...(-i D_2)\} ...(-i D_4)\}
	\end{eqnarray}
	Here $ (-i D_i)$ is encountered $\alpha_i$ times. For each $\alpha$  $o_{\alpha}(x)$ is a matrix - valued function of coordinates. This is matrix in internal space, but it does not contain gauge field $A$.

We also denote  $\hat Q:=Q(\hat x,\hat \pi)$. Then the following relationship holds:
\begin{equation}
	Q_{\cal{W}}(x,p)=(\hat Q)_{{W}}(x,p)\Big|_{A=0}=Q(x,p)  \label{exact}
\end{equation} 
Here $Q(x,p)$ is understood as
	\begin{eqnarray}
		Q(x,p)&=&\sum_{|\alpha| \le m} o_{\alpha}(x)  p^\alpha
	\end{eqnarray}
	that is to calculate $Q(x,p)$ we simply substitute to Eq. (\ref{sym0}) $p$ instead of $-i D$, and replace $\circ$ by ordinary product in order to obtain the above expression.  Therefore, we are able to substitute $Q_{\cal W}(x,p)$ by $Q(x,p)$.

The two point Green function $\hat{G} = - \langle \hat{\rho} \rangle$ is inverse to operator $\hat{Q}$. Its covariant Weyl symbol obeys the generalized Groenewold equation
\begin{equation}
	Q \bigstar G_W = 1
\end{equation}
Its solution may be expanded in powers of the covariant derivative $D_z$: 
\begin{equation}
	G_{{W}}(x,p)=\sum_{n\geq 0} G^{(n)}(x,p,z)|_{z=x}.
\end{equation}
Here $G^{(n)}(x,p,z)$ contains $n$ powers of derivative $D_z$. To the lowest order the generalized Groenewold equation becomes
\begin{equation}
	Q(x,p)\star G^{(0)}(x,p,z)=1 \label{bumb}
\end{equation}
where $\star = e^{\frac{i}{2}(\overleftarrow{\partial}_x\overrightarrow{\partial}_p-\overleftarrow{\partial}_p\overrightarrow{\partial}_x)}$ is the ordinary Moyal product.
In the next order we have $G^{(1)}=0$ and
\begin{align}
	G^{(2)}(x,p,z)&=-\frac{i}{2}G^{(0)}(x,p)\star\partial_{p_{\mu}}Q(x,p)\star \partial_{p_{\nu}}G^{(0)}(x,p)F_{\mu\nu}(z),\label{G2}
\end{align}
where $F_{\mu\nu}$ is the gauge field strength.




The Abelian vector current is defined as the Noether current corresponding to the gauge transformation 
\begin{align}
	\psi(x)\to e^{i\alpha(x)}\psi(x) \label{this} \\
	\bar\psi(x)\to\bar\psi(x)e^{-i\alpha(x)} \label{this1} 
\end{align}
where $\alpha(x) \in {\bold R}$. 

In terms of the covariant Wigner transform of the Green function, $G_{\cal W}$, the average vector current is given by 
\begin{equation}
	\langle J_{\mu}(x)\rangle=-\text{tr}_{D}\int\frac{d^4p}{(2\pi)^4}G_{\cal W}\partial_{p_\mu} Q \label{unrahv}
\end{equation}

{\it In the following for brevity we omit subscript $\cal W$  for  $G_{\cal W}$. }

\section{Noether procedure in the presence of additional symmetry  }

Let us consider the situation, when Dirac operator $\hat{Q}$ commutes  with Hermitian  matrix $\hat{T}$. Correspondingly, action in Eq. (\ref{ac}) is symmetric under the transformation
\begin{align}
	\psi(x)\to e^{i\alpha(x) \hat{T}}\psi(x) \nonumber \\
	\bar\psi(x)\to\bar\psi(x)e^{-i\alpha(x)\hat{T}} \label{this2} 
\end{align}
where $\alpha(x) \in {\bold R}$ does not depend on coordinates. The divergence  of the corresponding Noether current appears as response of the action to the above transformation with varying $\alpha(x)$.

The variation of the action under an infinitesimal version of \eqref{this2} is
\begin{align}
	\delta S&=-\text{tr}_{ D}\text{tr}_{H}\left(\delta\hat Q\hat\rho\right)\label{fade}\\
	\delta \hat Q&=[\hat Q,i\alpha(\hat x)\hat{T}] \label{rundmc}
\end{align}
Using the trace this can be written as
\begin{align}
	\delta S=\text{tr}_{ D}\text{tr}_{H}\left(i\alpha(\hat x)\hat{T}[\hat Q,\hat\rho]\right) \label{fader}
\end{align}
Using the fact that $\text{tr}_{H}(\hat X\hat Y)=\text{tr}_{\Gamma}(X_{\cal W}Y_{\cal W})$ (where $\text{tr}_\Gamma\equiv\int d^4x\int(2\pi)^{-4}d^4p$ is the trace w.r.t. the phase space) we get
\begin{align}
	\delta S&=\int d^4x~\alpha(x)\Gamma(x)\nonumber\\
	\Gamma(x)&=i\text{tr}_{D}\hat{T}\int\frac{d^4p}{(2\pi)^{4}}(Q\bigstar\rho_{\cal W}-\rho_{\cal W}\bigstar Q) \label{ug}
\end{align}
Using definition of the star and integrating over $p$ we get
\begin{align}
	\begin{split}
		\Gamma(x)&=i\text{tr}_{D}\hat{T}\int(2\pi)^{-8}d^4yd^4kd^4k'e^{-iy(k-k')}\nonumber\\
		&\Bigl(Q(x+y/2,k)\rho_{\cal W}(x+y/2,k')\\&-\rho_{\cal W}(x+y/2,k)Q(x+y/2,k')\Bigr)
		 \label{geo}
	\end{split}
\end{align}
We can write it as
\begin{align}
	\Gamma(x)&=i\text{tr}_{D}\hat{T}\int(2\pi)^{-8}d^4yd^4kd^4k'e^{-iy(k-k')}e^{y{\nabla_x}/2}\nonumber\\&\left(Q(x,k)\rho_{\cal W}(x,k')-\rho_{\cal W}(x,k)Q(x,k')\right) 
\end{align}
Integrating by parts we get
\begin{align}
	\Gamma(x)&=i\text{tr}_{D}\hat{T}\int(2\pi)^{-8}d^4yd^4kd^4k'e^{-iy(k-k')}\Bigl(e^{-i\partial_{k}{\nabla_x}/2}\nonumber\\&\left(Q(x,k)\rho_{\cal W}(x,k')\right)-e^{i\partial_{k'}{\nabla_x }/2}\left(\rho_{\cal W}(x,k)Q(x,k')\right)\Bigr) \label{her}
\end{align}
Expanding in powers of derivatives 
we get
\begin{align}
	\Gamma&=i\text{tr}_{D}\hat{T}\int\frac{d^4p}{(2\pi)^4}[Q,\rho_{\cal W}]+{\mathcal{\nabla}}_\mu j^{(T)}_\mu \\
	j^{(T)}_{\mu}&=\frac{1}{2}\text{tr}_{D}\hat{T}\int\frac{d^4p}{(2\pi)^4}\left(\partial_{p_\mu}Q\rho_{\cal W}+\rho_{\cal W}\partial_{p_\mu}Q\right)+...
\end{align}
Here $j$ is the miscroscopic current.
Dots mean the terms with higher powers of $\nabla_x$. When we average the microscopic current over the system sample these terms are irrelevant. Therefore, we define the quantum average of the {\it macroscopic} current $J^{(T)}$ as  
\begin{equation}
	\langle J^{(T)}_{\mu}\rangle=-\frac{1}{2}\text{tr}_{D}\int\frac{d^4p}{(2\pi)^4}G_{\cal W}\partial_{p_\mu}\{Q,\hat{T}\}  \label{unrah}
\end{equation}
If the symmetry generated by matrix $\hat T$ is not anomalous, the corresponding current is conserved. This follows from the invariance of partition function with respect to the transformation of Eq. (\ref{this2}). In the presence of anomaly (say, if $\hat{T}$ generates chiral symmetry) the divergence of the current is nonzero. 

Ordinary electric current is the particular case of the current protected by symmetry with $\hat{T} = 1$.

\section{Response of a macroscopic current to external field strength and chemical potential}

We substitute Eq. (\ref{G2}) to expression for the current of Eq. (\ref{unrah}), and obtain 
\begin{align}
	\begin{split}
		J^{(T)}_\mu&=S^{(T)}_{\mu}(x)+M^{(T)}_{\mu\alpha\beta}(x)\text(F_{\alpha\beta}) \label{bruha}
	\end{split} \\
	S^{(T)}_{\mu}(x)&=-\frac{1}{2}\int \frac{dp}{(2\pi)^4}\text{tr}_{D}\left(\{\hat{T},\partial_{p^\mu}Q\} G^{(0)}\right)\\
	M^{(T)}_{\mu\alpha\beta}(x)&=\frac{i}{8}\int \frac{dp}{(2\pi)^4}\text{tr}_{D}\Bigl( \{\hat{T},\partial_{p^\mu}Q\} G^{(0)}\nonumber\\&\star\partial_{p^{[\alpha}}Q\star \partial_{p^{\beta]}}G^{(0)}\Bigr)\label{M123}
\end{align}
The second term in Eq. (\ref{bruha})  defines response of the given current to external electromagnetic field. In the case when the latter is due to magnetic field only we can calculate the derivative of this response to chemical potential as 
\begin{equation}
	\frac{\partial}{\partial \mu} J^{(T,2)}_{i} \equiv \frac{\partial}{\partial \mu} M^{(T)}_{ijk}\epsilon_{jkl} B_l
\end{equation}
Chemical potential appears as the shift of Matsubara frequency $p_4$ : $ i p_4 \to ip_4 + \mu$. Therefore, 
\begin{align}
	\frac{\partial }{\partial \mu}M^{(T)}_{ijk}(x)&=\frac{1}{8}\int \frac{dp}{(2\pi)^4}\frac{\partial}{\partial p_4}\text{tr}_{D}\Bigl( \{\hat{T},\partial_{p^i}Q\} G^{(0)}\nonumber\\&\star\partial_{p^{[j}}Q\star \partial_{p^{k]}}G^{(0)}\Bigr)\label{M123mu}
\end{align}
Due to the presence of poles in the expression standing inside the integral, the integral is not zero, but it is reduced to integrals over the hyperplanes $\Sigma_{p_4}$ for  $p_4 = \pm \epsilon, \pm \infty$, $\epsilon \to 0$: 
\begin{align}
	&\frac{\partial }{\partial \mu}M^{(T)}_{ijk}(x)=-\frac{1}{8}\sum_{p_4 = \pm \epsilon, \pm \infty}\int_{\Sigma_{p_4}} \frac{dp_1 dp_2 dp_3}{(2\pi)^4}\nonumber\\&\text{tr}_{D}\left(\{\hat{T}, \partial_{p^i}Q\} G^{(0)}\star\partial_{p^{[j}}Q\star \partial_{p^{k]}}G^{(0)}\right)\label{M123mu}
\end{align}
Ultraviolet regularization removes from this sum the contribution of $p_4 = \pm \infty$. For example, lattice regularization makes integrals over $p_4$ along a circle instead of $\bold R$, while Pauli - Villars regularization adds the contributions of Pauli - Villars regultors, which cancel the contribution of $p_4 = \pm \infty$ in Eq. (\ref{M123mu}). 

If symmetry generated by matrix $\hat T$ is anomalous, then the ultraviolet regularization breaks it, and $\{\hat{T},\partial_p Q\} \ne 2 \hat{T}\partial_p Q$ for all values of momenta. However, in a vicinity of physical singularities that mark positions of the Fermi points we still can use  equality $\{\hat{T},\partial_p Q\} = 2 \hat{T}\partial_p Q$.

The remaining contributions to Eq. (\ref{M123mu}) can be reduced to integrals over {the  hypersurfaces $p_4 = \pm \epsilon$} that enclose the poles of the expression standing inside the integral: 
\begin{equation}
	\frac{\partial}{\partial \mu}\langle J^{(T,2)}_i\rangle = \sigma^{(T)} B_i.
\end{equation}
Here {$B_i=\frac{1}{2}\epsilon_{ijk}F_{jk}$ is the magnetic field}, averaging on the left hand side is over the whole system volume, while 
\begin{equation}
	\sigma^{(T)} = \frac{{ N}^{(T)}_3}{4\pi^2} \label{sigmaCSE}
\end{equation}
and
{\begin{eqnarray}
		N^{(T)}_3&=&{-}\frac{1}{V \,24\pi^2}\sum_{p_4 = \pm \epsilon}\int d^3 x\int_{\Sigma_{p_4}} \nonumber \\ && \text{tr}_{D}\left(\hat{T}G^{(0)}\star dQ\star \wedge dG^{(0)}\star\wedge dQ \right)\label{N3CSE}
\end{eqnarray}}
\mz{(Hyperplanes $\Sigma_{\pm \epsilon}$ have opposite orientations.)}
The  above expression is a topological invariant due to the presence of integration over the three - dimensional momenta $\bar{p}$ and coordinates $\bar{x}$.

There exists also the representation of Eq. (\ref{N3CSE}) in the form of an integral over the \mz{hypersurface of more general form} embracing the singularities of expression standing in the integral in Eq. (\ref{N3CSE}). 

In particular, we can represent \begin{eqnarray} 
				N^{(T)}_3	&=&{-}\frac{1}{V \,24\pi^2}\int d^3 x\int_{\Sigma(\bar{x})} \text{tr}_{D}\Bigl(\hat{T}\tilde{G}^{(0)}\star d\tilde{Q}\nonumber\\&&\star \wedge d\tilde{G}^{(0)}\star\wedge d\tilde{Q} \Bigr) \label{NT3}
\end{eqnarray}
			In transition from Eq. (\ref{N3CSE}) to this expression we perform a smooth modification of Dirac operator: $$Q(x,(\bar{p},\pm \epsilon)) \to Q(x,(\bar{p},\pm \omega_\pm(\bar{p},\bar{x}))) \equiv \tilde{Q}(\bar{x},\bar{p}).$$ Here $\omega_\pm(\bar{p},\bar{x})$ is a function, which describes the dependence of $p_4$ on $\bar{p}$ on the hypersurface $\Sigma(\bar{x})$ . The latter hypersurface in momentum space for any $\bar{x}$ embraces the singularities of the expression standing inside the integral. For the case of a homogeneous system these singularities appear along the Fermi surface (or Fermi point). In the inhomogeneous case the geometrical place in momentum space of such singularities depends on $\bar{x}$ and generalizes the notion of the Fermi surface. We will refer to these surfaces in momentum space as to the coordinate depending Fermi surfaces. 
			In turn, in transition from Eq. (\ref{N3CSE}) we deform also the Green function as $G^{(0)}(x,(\bar{p},\pm \epsilon)) \to \tilde{G}^{}(\bar{x},\bar{p})$, where the latter function obeys the {\it transformed} Groenewold equation 
			\begin{equation}
			\tilde{Q}^{}(\bar{x},\bar{p})e^{\frac{i}{2}(\overleftarrow{\partial}_{\bar{x}}\overrightarrow{\partial}_{\bar{p}} - \overleftarrow{\partial}_{\bar{p}}\overrightarrow{\partial}_{\bar{x}})} \tilde{G}^{(0)}(\bar{p},\bar{x})=1\label{Groenewold}
			\end{equation}
			The given smooth modification $Q \to \tilde{Q}$ and $G^{(0)}\to \tilde{G}^{(0)}$ does not result in a jump of the value of $N_3$ because defined this way it is the topological invariant. \mz{However, we should require that matrix $\hat T$ commutes with $\tilde{G}^{(0)}(\bar{x},\bar{p})$ and $\tilde{Q}^{}(\bar{x},\bar{p})$.  }  
	  
\mz{Dirac operator ${Q}$ is regular, but the Groenewold equation Eq. (\ref{bumb}) determines ${G}^{(0)}$, which may have singularities. We assume that all these singularities are situated at $p_4 = 0$.  Notice that the star product contains only derivatives with respect to spatial components of momenta. Therefore, we can calculate expression standing inside the integral in Eq. (\ref{N3CSE}) within each hypersurface $p_4 = const$ separately. As a result it has a discontinuity accross the hyperplane $p_4=0$ because  the value of ${G}^{(0)}$ just below the singularity differs from its value just above it. Suppose that at each $x$ the distance in momentum space between the disconnected pieces of Fermi surface is much larger than the inverse correlation length. Then there exists the region between these pieces, where there is no jump in ${G}^{(0)}$ accross the hyperplane $p_4=0$, and there in Eq. (\ref{N3CSE}) the pieces of $\Sigma_\epsilon$ and $\Sigma_{-\epsilon}$ cancel each other at $\epsilon \to 0$.  Then for any $\bar{x}$ the hypersurface $\Sigma = \Sigma_\epsilon \bigcup  \Sigma_{-\epsilon}  = \bigcup_{i} \Sigma^{(i)} $ at $\epsilon \to 0$ may be divided into  the hypersurfaces $\Sigma^{(i)} $ encompassing  the disconnected Fermi surfaces/Fermi points. }

In the presence of proper ultraviolet regularization the Bloch theorem (absence of total electric current in equilibrium) in line with the Nielsen - Ninomiya theorem (absence of a chiral fermion in lattice regularization) gives rise to the conclusion that	$N^{(T)}_3 = 0$ for $\hat{T} = 1$. However, the values of  	
\begin{eqnarray} 
	N^{(T,i)}_3&=&{-}\frac{1}{V \,24\pi^2}\int d^3 x\int_{\Sigma^{(i)}} \text{tr}_{D}\Bigl(\hat{T}{G}^{(0)}\star d{Q}\nonumber\\&&\star \wedge d{G}^{(0)}\star\wedge d{Q} \Bigr) \label{Ni}
\end{eqnarray}
may remain nonzero not only for $\hat{T} \ne 1$. If tuning the Fermi level we are able to reduce  the $i$ - th Fermi surface to a point, we are speaking of the ($\bar{x}$ depending) Fermi point. 

In this case Eq. (\ref{Ni}) is a topological invariant (protected by symmetry $\hat{T}$) responsible for the stability of the Fermi point. Modifying the system smoothly one cannot change its value until the two Fermi points/Fermi surfaces meet each other and annihilate. The latter may occur if the two corresponding values of $N^{(T,i)}_3$ are equal in absolute values and opposite in signs. 

\section{Arbitrary parametrization of hypersurface $\Sigma^{}$.}

It is instructive to represent Eq. (\ref{NT3}) in an arbitrary parametrization of the hypersurface $\Sigma^{}$ by real numbers $\bar{k} = (k_1,k_2,k_3)$. We still assume that the positions of Fermi surfaces/Fermi points depend on $\bar{x}$. However, we suppose that for any $\bar{x}$ exists the common hypersurface in momentum space that does not depend on $\bar{x}$ but surrounds the given Fermi surface for {\it any} $\bar{x}$. This means that the corresponding function $\omega_\pm(\bar{p})$ depends only on $\bar{p}$. We express $\bar{p}(\bar{k})$ through the new parameters $\bar{k}$. In terms of the latter the necessary   representation reads 
\begin{align}
	{N}_3^{(T)}
	&=\frac{\epsilon_{ijk}}{24 \pi^2 }
	\int_{\Sigma^{}}d^3 k
	\int   \frac{d^3 \xi}{{ V}}
	\tr \Bigl[\hat{T}
	\tilde{G}\circ \partial_{k_i} \tilde{ Q} \circ\nonumber\\& \tilde{ G}
	\circ \partial_{k_j} \tilde{ Q}\circ \tilde{ G} \circ  \partial_{k_k} \tilde{ Q}
	\Bigr]\label{Ncompl0}
\end{align}
with 
\begin{align}
	\circ &= e^{\frac{i}{2}\Bigl(\overleftarrow{\partial_{\xi^i}}\overrightarrow{\partial_{k_i}} - \overleftarrow{\partial_{k_i}}\overrightarrow{\partial_{\xi^i}} \Bigr)}\label{eqo}
\end{align}
and  $
{ V} = \int d^3 \xi
$.
Correspondingly, the limit of infinitely large $ V$ in Eq. (\ref{Ncompl0})  is to be considered. 
Vector $\bar{\xi}$ represents the new parametrization of coordinate space. These new coordinates are determined by equation 
$
\frac{\partial \xi^i(\bar{k},\bar{x})}{\partial x^j} = \frac{\partial p_j(\bar{k})}{\partial k_i}
$.
Therefore, 
$
\xi^i =\frac{\partial p_j(\bar{k})}{\partial k_i} x^j \, \quad (i,j = 1,2,3)
$.
Correspondingly, $\int d^3 \xi = {\rm Det}\,\frac{\partial p(\bar{k})}{\partial k}\, \int d^3 x$. 

By $\tilde Q$ in Eq. (\ref{Ncompl0}) we understand 
\begin{align}
	\tilde{Q}(\bar{p}(\bar{k}),\bar{x}(\bar{k},\bar{\xi})) \equiv  Q((\bar{p}(\bar{k}),\pm \omega_\pm(\bar{p}(\bar{k}))),x(\bar{k},\bar{\xi})) \label{B}
\end{align}
with $x^j = (\frac{\partial k_k}{\partial p_j} \, \xi^k)$, ($i,j = 1,2,3$).
Here there is no dependence on $x^4$ because we consider the equilibrium system, while 
$p = (\bar{p}(\bar{k}), \pm \omega_\pm(\bar{k}))$.

By $\tilde{G}$ we then understand the function inverse to $\tilde{Q}$ with respect to the $\circ$ product: 
\begin{align}
	\tilde{Q} \circ \tilde{G} = 1\label{C}
\end{align}

One can see that Eq. (\ref{Ncompl0}) is reparametrization - invariant. Namely, it is invariant under the  transformation 
\begin{align}
	{k}_i \to \tilde{k}_i({k}), \quad {\xi} \to \tilde{\xi}^j = \frac{\partial {k}_i}{\partial \tilde{k}_j} {\xi}^i  
\end{align}

One can check easily that the $\circ$ product of Eq. (\ref{eqo}) is equivalent to the $\star$ product of Eq. (\ref{Groenewold}). Transition between the two is given by the reparamtetrization transformation $x \to \xi, p \to k$. Recall that in Eq. (\ref{Groenewold}) the differentiation with respect to $\bar{p}$ entering the Moyal product takes into account the dependence of $p_4$  on $\bar{p}$. The particular case when $\bar{\xi}$ coincides with $\bar{x}$ while $\bar{k}$ coincides with $\bar{p}$ brings us back to Eq. (\ref{NT3}) (in the particular case when $\Sigma^{}$ does not depend on $x$).

\section{Toy models with nontrivial $N_3^{(T)}$. }

\subsection{Basic model. Calculation of $N_3$}

We will build our toy model in such a way that it resembles somehow the system of fermions in the presence of constant magnetic field.  We summarize here the results. The detailed calculation is presented in Supplementary materials.
Let us consider the system with Dirac operator of the form
\begin{equation}
	\hat{Q} = i \kappa \hat{p}_4 \hat{p}_3 - \hat{p}_3 -  \frac{\Bigl(\hat{p}_1^2 + (\hat{p}_2 - \hat{x}_1 B)^2 \Bigr) }{2m} + \mu
\end{equation}
with parameters $\kappa$, $m$, $\mu$, and $B$. The model with the Dirac operator of this type may appear as  effective description of an interacting system (the interactions cause change of the term $p_4 \to \kappa p_4 p_3$). 
The associated quantum field theory is well defined, and the corresponding topological invariant is given by Eq. (\ref{Ni}). (Obviousely, in this case $\hat{T} = 1$.) In Supplementary materials we present calculation of $N_3$ for this model for the relatively arbitrary form of surface $\Sigma$ with 
 $p_4 =  \omega_\pm(p_3)$ (we assume that $\omega_+(p_3)>0$ while $\omega_-(p_3) < 0$):
$$
N_3^{} = N_3^{\omega_+} - N_3^{\omega_-}
$$ 
where 
\begin{eqnarray} 
	N^{\omega}_3&	= &{-}\frac{1}{V \,24\pi^2}\int d^3 x\int_{p_4 = \omega} \text{tr}_{D}\Bigl({G}_W\star d{Q}_W\nonumber\\&&\star \wedge d{G}_W\star\wedge d{Q}_W \Bigr) \label{Ni20}
\end{eqnarray}
This expression can be calculated using technique developed in \cite{zubkov2020topological}. We denote
\begin{equation}
	\hat{H} = \frac{\Bigl(\hat{p}_1^2 + (\hat{p}_2 - \hat{x}_1 B)^2 \Bigr) }{2m} - \mu, \quad \hat{H}|n,p_2\rangle = E_n |n,p_2\rangle \label{energylevels}
\end{equation}
and obtain 
\begin{equation}
	N_3^\omega = {\rm sign}\,\omega\,\sum_n \Theta( - E_n(B,m)),
\end{equation}
where $E_n$ is the energy of the $n$ - th Landau level 
\begin{equation}
	E_n(B,m)  = \frac{B}{m} (n+1/2) -\mu, \quad n = 0,1,...
\end{equation}
This gives
$$
N_3^{} = 2 \Bigl[\frac{\mu - B/(2m)}{B/m}\Bigr] \theta(\mu - B/(2m))
$$

{As it should there is no dependence of the result on the form of functions $\omega_\pm(p_3)$.}

The singularities of the Green function appear if the corresponding energy levels (chemical potential included) are vanishing. Let us define the wave functions 
\begin{eqnarray}
&& \langle x_1|n, p_2\rangle = \Psi_n(x - p_2/B)
\end{eqnarray}
We denote by numbers $n$ the Landau levels, while these levels are degenerate (corresponding to number $p_2$). We obtain for the Green function:
\begin{eqnarray}
	&&G_W(x_1,p_1,p_2,p_3,p_4) = \sum_n \int dy e^{ip_1 y}\nonumber\\&&\frac{\Psi_n(x_1 - y/2 - p_2/B) \bar{\Psi}_n(x_1 + y/2 - p_2/B)}{ip_4 p_3 \kappa - p_3 - \frac{B}{m}(n+1/2) + \mu}
\end{eqnarray}
If $\mu$ does not coincide with either of the levels $\frac{B}{m}(n+1/2)$, the singularities of the above expresion appear at 
\begin{equation}
	p_4 = 0, \quad p_3 = \mu - \frac{B}{m}(n+1/2), n = 0, 1, ...  
\end{equation}
These are the two dimensional hypersurfaces in the hyperplane $p_4 =0$ parametrized by $p_1,p_2$. Their positions do not depend on $x$.

Let us consider surface $\Sigma$ of the form of a collection of three - dimensional hypersurfaces $\Sigma^{(l)}$ surrounding the poles of the Green function.  The corresponding expression for the partial invariant is 
\begin{eqnarray}
	&&N_3^{(l)} = \Theta( - E_l(B,m)) +   \sum_n \Theta( - E_n(B,m)) = \label{Nie}\\&& =  \Theta(\mu - (2l+1) B/(2m)) +  \Bigl[\frac{\mu - B/(2m)}{B/m}\Bigr] \theta(\mu - B/(2m))\nonumber
\end{eqnarray}
(The detailed derivation is given in Supplementary materials.)

\subsection{Other variations of the model}

\subsubsection{A model with $M=1$}

We can modify the model considered above in several ways. First of all, let us consider the system with Dirac operator of the form
\begin{eqnarray}
	\hat{Q} &=& i \kappa \hat{p}_4 \Bigl[\frac{\hat{p}_1^2 + (\hat{p}_2 - \hat{x}_1 B)^2  }{2m} - \mu\Bigr] - \hat{p}_3 \nonumber\\&&-  \frac{\Bigl(\hat{p}_1^2 + (\hat{p}_2 - \hat{x}_1 B)^2 \Bigr) }{2m} + \mu
\end{eqnarray}
with parameters $\kappa$, $m$, $\mu$, and $B$. The following consideration repeats the one presented above, and we arrive at the expression for the topological invariant is the same  as for the above considered model 
$$
N_3^{} = 2\Bigl[\frac{\mu - B/(2m)}{B/m}\Bigr] \theta(\mu - B/(2m))
$$
Also the positions of the Fermi surfaces are the same as for the original model.
As above, we can also  define quantities $N_3^{(i)}$ that are given by integrals along the hypertubes surrounding the disconnected pieces of the Fermi surface. We arrive then at the same expression of Eq. (\ref{Nie}).

\subsubsection{A model with topological invariant protected by symmetry ($M=4$)}

Let us consider the matrix extension of the model with the Dirac operator of the form
\begin{eqnarray}
	&&\hat{Q} = i \kappa \hat{p}_4 \Bigl(\hat{p}_1 \gamma^0\gamma^1 + (\hat{p}_2 - \hat{x}_1 B) \gamma^0 \gamma^2+i \mu \gamma^0 \gamma^3 \Bigr)\nonumber\\&&- i\hat{p}_3\gamma^0 \gamma^3  - \hat{p}_1 \gamma^0\gamma^1 - (\hat{p}_2 - \hat{x}_1 B) \gamma^0 \gamma^2-i\mu \gamma^0 \gamma^3 
\end{eqnarray}
with parameters $\kappa$ and $B$ and ordinary gamma - matrices $\gamma^i$ taken in chiral representation.
 One may define the corresponding topological invariant protected by symmetry  given by Eq. (\ref{NT3}) with $\hat{T} = \gamma^5$. We define the auxiliary Hamiltonian
 \begin{equation}
 	\hat{H} =  i \sigma^3 \sigma^1 \hat{p}_1 + i \sigma^3 \sigma^2  (\hat{p}_2 - \hat{x}_1 B) - \mu, \quad \hat{H}|n,p_2\rangle = E_n |n,p_2\rangle,
 \end{equation}
 where $E_n$ is the energy of the $n$ - th Landau level 
\begin{equation}
	E_n(B,m)  = \pm \sqrt{2B n}-\mu, \quad n = 0,1,...
\end{equation}
The naively calculated expression for $N_3^{\gamma^5}$ is divergent in ultraviolet: 
\begin{equation}
	N_3^{\gamma^5} = 4\, \sum_n \Theta(E_n(B,\mu)),\label{N3g5}
\end{equation}
The situation with this divergency is similar to the one of the quantum Hall effect in graphene. Formally the corresponding expression obtained using the above mentioned machinery applied to the low energy continuum model is given by Eq. (\ref{N3g5}) multiplied by the inverse Klitzing constant. Of course, the experiment shows a different result. Namely, the QHE conductivity is vanishing at half filling ($\mu = 0$). 

The explanation for this puzzle is that the proper ultraviolet regularization subtracts the contribution of Landau levels with negative $E_n$. At the same time the half of the contribution of the LLL ($E_0 =0$) is subtracted. Technically the subtraction is achieved authomatically, when proper regularization is added. Say, lattice regularization modifies expression of Eq. (\ref{N3g5}) at large values of $n$: the corresponding Landau levels contribute with their Chern numbers that differ from $1$. In particular, at certain negative values of $n$ the contributions are large and negative, so that at $\mu = 0$ the resulting expression is precisely zero. The Pauli - Villars regularization gives much more transparent solution. Namely, the contributions of Pauli - Villars regulators with large mass cancel one by one all contributions to $N_3^{\gamma^5}$ of Landau levels with $E_n < 0$, while only half contribution of $E_0$ is cancelled. 

Thus we come to expression 
  \begin{equation}
  	N_3^{\gamma^5} = \Bigl(4\, \Bigl[\frac{\mu^2}{2B}\Bigr] + 2\Bigr)\,{\rm sign}\,\mu  \label{N3g5_}
  \end{equation} 

\subsubsection{Models with curved Fermi surfaces }

Let us consider another modification of the model, in which the Fermi surfaces are already not planes $p_3 = const$ but have cylindrical or spherical form. We start from the Dirac operator 
\begin{equation}
	\hat{Q} = i \kappa \lambda \hat{p}_4 (\hat{H}-\mu) - \hat{{p}}_r\gamma^5 - (\hat{H}  + \mu)\lambda 
\end{equation}
It is assumed that $\hat{H}$ commutes with $\gamma^5$. $\kappa$ and $\lambda$ are parameters. In order to define the Hamiltonian $\hat{H}$ we pass to the spherical (cylindrical) coordinates in momentum space: $(p_r, \theta,\phi)$ are  spherical coordinates ($(p_r, p_z,\phi)$ are  cylindrical  coordinates). In the first case we define new parametrization $\bar{k} = (p_r,p_\theta,p_\phi)$ of momentum space $(p_r\in (0,\infty), p_\theta =  \theta/b \in (0,  \pi/b), p_\phi =\phi/a \in (-\pi/a,  \pi/a)$. In the second case $\bar{k} = (p_r,p_z,p_\phi)$ with $(p_r\in (0,\infty), p_z \in (-\pi/b,  \pi/b), p_\phi =\phi/a \in (-\pi/a,  \pi/a)$ 

We assume that parameters $b,a \to \infty$, and in both cases define
\begin{eqnarray}
	\hat{H}& =& f_a(k_3 +  B \hat{\xi}_2)\gamma^0\gamma^1 +  f_b(k_2) \gamma^0\gamma^2  
\end{eqnarray}
with $f_a(p) = \frac{1}{a}{\rm sin}\,pa + \frac{1}{a}(1-{\rm cos} \,pa)$, which gives
\begin{eqnarray}
	\hat{H}& \approx & (k_3 +  B \hat{\xi}_2)\gamma^0\gamma^1 +  k_2 \gamma^0\gamma^2  
\end{eqnarray}
with $\hat{\xi}_i = i \partial_{k_i}$.

Then $p_r$ commutes with the Hamiltonian, and spectrum of this Hamiltonian for any $p_r$ is: 
\begin{equation}
	\hat{H} |n, k\rangle = {\cal E}_n |n, k\rangle
\end{equation}
with ${\cal E}_n =  \sqrt{2B|n|} \,{\rm sign}\,n$. 
Each energy level is degenerate, the degenerate states are enumerated by integer number $k = 0, ...$ and the chirality (left - handed/right - handed). 
We also denote $E_n = \lambda({\cal E}_n - \mu)$ 
 and  arrive at 
$$
N_3^{} = 2 \sum_{n} \Theta(\mu - \sqrt{2 B |n|} \, {\rm sign}\, n)
$$
As in the previous example, this expression is divergent, but being regularized, it gives
$$
N_3^{} = 2 \, {\rm sign}\,(\mu)\, \Bigl(1/2+\sum_{n>0} \Theta(\mu - \sqrt{2 B |n|} )\Bigr)
$$

In this case the Fermi surfaces have the form of concentric spheres (or cylinders with axis z) with radii 
\begin{equation}
	p_4 = 0, \quad p_r = \mp \lambda(\mu - {\rm sign}\,(n)\, \sqrt{2 B |n|}), n = 0, \pm 1, ...  
\end{equation}
(Only those values of $n$ contribute, for which the right hand side of this expression is not negative.)
The particular interesting case corresponds to $\mu = \delta$, $\delta \to 0$. Then the only Fermi surface shrinks to a very small sphere (or cylinder). The corresponding invariant $N_3^{(0)} = {\rm sign }\, \delta$ protects this Fermi surface.

\section{Conclusions}
To conclude, in the present paper we discussed the situation, when a fermionic quantum field system is not homogeneous. Then the coordinate depending Fermi surface can be defined as the position in momentum space of the singularities of the Wigner transformed Green function. If the Fermi surface defined in this way consists of several pieces not connected one to another, the topological stability of these pieces is protected by the topological invariants considered in the present paper. The important property of these invairants is that they are expressed through the Wigner transformed Green functions, and contain Moyal products. The case, when the given system is homotopic to a homogeneous one brings us back to the previously considered topological invariants \cite{Volovik2003a} of Eqs. (\ref{cal11}) and (\ref{cal12}). However, the nontrivial cases exist, when the given topological invariants are nonzero due to the presence of Moyal products rather than due to the matrix structure of Green function. We consider the corresponding toy models with various forms of Fermi surfaces, and various matrix structures. It is worth mentioning that the particular cases exist of the given invariants, in which they are responsible for the non - dissipative transport phenomena. In particular, the considered invariant protected by chiral symmetry enters expression for the conductivity of chiral separation effect \cite{XZ2024}.   

\section*{Funding}

This work was supported by ongoing institutional funding. No additional grants to carry out or direct this particular research were obtained.

\section*{Conflict of interests}

The author of this work declares that he has no conflicts of interest.

\bibliography{wigner3,cross-ref,biblio_corrected} 

\begin{thebibliography}{29}%
\makeatletter
\providecommand \@ifxundefined [1]{%
 \@ifx{#1\undefined}
}%
\providecommand \@ifnum [1]{%
 \ifnum #1\expandafter \@firstoftwo
 \else \expandafter \@secondoftwo
 \fi
}%
\providecommand \@ifx [1]{%
 \ifx #1\expandafter \@firstoftwo
 \else \expandafter \@secondoftwo
 \fi
}%
\providecommand \natexlab [1]{#1}%
\providecommand \enquote  [1]{``#1''}%
\providecommand \bibnamefont  [1]{#1}%
\providecommand \bibfnamefont [1]{#1}%
\providecommand \citenamefont [1]{#1}%
\providecommand \href@noop [0]{\@secondoftwo}%
\providecommand \href [0]{\begingroup \@sanitize@url \@href}%
\providecommand \@href[1]{\@@startlink{#1}\@@href}%
\providecommand \@@href[1]{\endgroup#1\@@endlink}%
\providecommand \@sanitize@url [0]{\catcode `\\12\catcode `\$12\catcode
  `\&12\catcode `\#12\catcode `\^12\catcode `\_12\catcode `\%12\relax}%
\providecommand \@@startlink[1]{}%
\providecommand \@@endlink[0]{}%
\providecommand \url  [0]{\begingroup\@sanitize@url \@url }%
\providecommand \@url [1]{\endgroup\@href {#1}{\urlprefix }}%
\providecommand \urlprefix  [0]{URL }%
\providecommand \Eprint [0]{\href }%
\providecommand \doibase [0]{https://doi.org/}%
\providecommand \selectlanguage [0]{\@gobble}%
\providecommand \bibinfo  [0]{\@secondoftwo}%
\providecommand \bibfield  [0]{\@secondoftwo}%
\providecommand \translation [1]{[#1]}%
\providecommand \BibitemOpen [0]{}%
\providecommand \bibitemStop [0]{}%
\providecommand \bibitemNoStop [0]{.\EOS\space}%
\providecommand \EOS [0]{\spacefactor3000\relax}%
\providecommand \BibitemShut  [1]{\csname bibitem#1\endcsname}%
\let\auto@bib@innerbib\@empty
\bibitem [{\citenamefont {Thouless}\ \emph {et~al.}(1982)\citenamefont
  {Thouless}, \citenamefont {Kohmoto}, \citenamefont {Nightingale},\ and\
  \citenamefont {den Nijs}}]{Thouless1982}%
  \BibitemOpen
  \bibfield  {author} {\bibinfo {author} {\bibfnamefont {D.~J.}\ \bibnamefont
  {Thouless}}, \bibinfo {author} {\bibfnamefont {M.}~\bibnamefont {Kohmoto}},
  \bibinfo {author} {\bibfnamefont {M.~P.}\ \bibnamefont {Nightingale}},\ and\
  \bibinfo {author} {\bibfnamefont {M.}~\bibnamefont {den Nijs}},\ }\bibfield
  {title} {\bibinfo {title} {{Quantized Hall Conductance in a Two-Dimensional
  Periodic Potential}},\ }\href {https://doi.org/10.1103/PhysRevLett.49.405}
  {\bibfield  {journal} {\bibinfo  {journal} {Phys. Rev. Lett.}\ }\textbf
  {\bibinfo {volume} {49}},\ \bibinfo {pages} {405} (\bibinfo {year}
  {1982})}\BibitemShut {NoStop}%
\bibitem [{\citenamefont {Kaufmann}\ \emph {et~al.}(2016)\citenamefont
  {Kaufmann}, \citenamefont {Li},\ and\ \citenamefont
  {Wehefritz-Kaufmann}}]{Kaufmann:2015lga}%
  \BibitemOpen
  \bibfield  {author} {\bibinfo {author} {\bibfnamefont {R.~M.}\ \bibnamefont
  {Kaufmann}}, \bibinfo {author} {\bibfnamefont {D.}~\bibnamefont {Li}},\ and\
  \bibinfo {author} {\bibfnamefont {B.}~\bibnamefont {Wehefritz-Kaufmann}},\
  }\bibfield  {title} {\bibinfo {title} {{Notes on topological insulators}},\
  }\href {https://doi.org/10.1142/S0129055X1630003X} {\bibfield  {journal}
  {\bibinfo  {journal} {Rev. Math. Phys.}\ }\textbf {\bibinfo {volume} {28}},\
  \bibinfo {pages} {1630003} (\bibinfo {year} {2016})},\ \Eprint
  {https://arxiv.org/abs/1501.02874} {arXiv:1501.02874 [math-ph]} \BibitemShut
  {NoStop}%
\bibitem [{\citenamefont {Avron}\ \emph {et~al.}(1983)\citenamefont {Avron},
  \citenamefont {Seiler},\ and\ \citenamefont {Simon}}]{Avron1983}%
  \BibitemOpen
  \bibfield  {author} {\bibinfo {author} {\bibfnamefont {J.~E.}\ \bibnamefont
  {Avron}}, \bibinfo {author} {\bibfnamefont {R.}~\bibnamefont {Seiler}},\ and\
  \bibinfo {author} {\bibfnamefont {B.}~\bibnamefont {Simon}},\ }\bibfield
  {title} {\bibinfo {title} {Homotopy and quantization in condensed matter
  physics},\ }\href {https://doi.org/10.1103/PhysRevLett.51.51} {\bibfield
  {journal} {\bibinfo  {journal} {Phys. Rev. Lett.}\ }\textbf {\bibinfo
  {volume} {51}},\ \bibinfo {pages} {51} (\bibinfo {year} {1983})}\BibitemShut
  {NoStop}%
\bibitem [{\citenamefont {Fradkin}(1991)}]{Fradkin1991}%
  \BibitemOpen
  \bibfield  {author} {\bibinfo {author} {\bibfnamefont {E.}~\bibnamefont
  {Fradkin}},\ }\href@noop {} {\emph {\bibinfo {title} {{Field Theories of
  Condensed Matter Physics}}}}\ (\bibinfo  {publisher} {Addison Wesley
  Publishing Company, New York},\ \bibinfo {year} {1991})\BibitemShut {NoStop}%
\bibitem [{\citenamefont {Tong}(2016)}]{Tong:2016kpv}%
  \BibitemOpen
  \bibfield  {author} {\bibinfo {author} {\bibfnamefont {D.}~\bibnamefont
  {Tong}},\ }\href@noop {} {\bibinfo {title} {Lectures on the quantum hall
  effect}} (\bibinfo {year} {2016}),\ \Eprint
  {https://arxiv.org/abs/arXiv:1606.06687} {arXiv:1606.06687 [hep-th]}
  \BibitemShut {NoStop}%
\bibitem [{\citenamefont {Hatsugai}(1997)}]{Hatsugai1997}%
  \BibitemOpen
  \bibfield  {author} {\bibinfo {author} {\bibfnamefont {Y.}~\bibnamefont
  {Hatsugai}},\ }\bibfield  {title} {\bibinfo {title} {{Topological aspects of
  the quantum Hall effect}},\ }\href@noop {} {\bibfield  {journal} {\bibinfo
  {journal} {J. Phys. Condens. Matter}\ }\textbf {\bibinfo {volume} {9}},\
  \bibinfo {pages} {2507} (\bibinfo {year} {1997})}\BibitemShut {NoStop}%
\bibitem [{\citenamefont {Qi}\ \emph {et~al.}(2008)\citenamefont {Qi},
  \citenamefont {Hughes},\ and\ \citenamefont {Zhang}}]{Qi2008}%
  \BibitemOpen
  \bibfield  {author} {\bibinfo {author} {\bibfnamefont {X.-L.}\ \bibnamefont
  {Qi}}, \bibinfo {author} {\bibfnamefont {T.~L.}\ \bibnamefont {Hughes}},\
  and\ \bibinfo {author} {\bibfnamefont {S.-C.}\ \bibnamefont {Zhang}},\
  }\bibfield  {title} {\bibinfo {title} {Topological field theory of
  time-reversal invariant insulators},\ }\href
  {https://doi.org/10.1103/PhysRevB.78.195424} {\bibfield  {journal} {\bibinfo
  {journal} {Phys. Rev. B}\ }\textbf {\bibinfo {volume} {78}},\ \bibinfo
  {pages} {195424} (\bibinfo {year} {2008})}\BibitemShut {NoStop}%
\bibitem [{\citenamefont {Ishikawa}\ and\ \citenamefont
  {Matsuyama}(1986)}]{IshikawaMatsuyama1986}%
  \BibitemOpen
  \bibfield  {author} {\bibinfo {author} {\bibfnamefont {K.}~\bibnamefont
  {Ishikawa}}\ and\ \bibinfo {author} {\bibfnamefont {T.}~\bibnamefont
  {Matsuyama}},\ }\bibfield  {title} {\bibinfo {title} {{Magnetic field induced
  multi component QED in three-dimensions and quantum Hall effect}},\
  }\href@noop {} {\bibfield  {journal} {\bibinfo  {journal} {Z. Phys. C}\
  }\textbf {\bibinfo {volume} {33}},\ \bibinfo {pages} {41} (\bibinfo {year}
  {1986})}\BibitemShut {NoStop}%
\bibitem [{\citenamefont {Volovik}(1988)}]{Volovik1988}%
  \BibitemOpen
  \bibfield  {author} {\bibinfo {author} {\bibfnamefont {G.~E.}\ \bibnamefont
  {Volovik}},\ }\bibfield  {title} {\bibinfo {title} {{An analog of the quantum
  Hall effect in a superfluid 3He film}},\ }\href@noop {} {\bibfield  {journal}
  {\bibinfo  {journal} {JETP}\ }\textbf {\bibinfo {volume} {67}},\ \bibinfo
  {pages} {9} (\bibinfo {year} {1988})},\ \bibinfo {note} {zhETF, Vol. 94, No.
  3(9), 123 (1988)}\BibitemShut {NoStop}%
\bibitem [{\citenamefont {Volovik}(2003)}]{Volovik2003a}%
  \BibitemOpen
  \bibfield  {author} {\bibinfo {author} {\bibfnamefont {G.~E.}\ \bibnamefont
  {Volovik}},\ }\href@noop {} {\emph {\bibinfo {title} {{The Universe in a
  Helium Droplet}}}}\ (\bibinfo  {publisher} {Clarendon Press},\ \bibinfo
  {address} {Oxford},\ \bibinfo {year} {2003})\BibitemShut {NoStop}%
\bibitem [{\citenamefont {Coleman}\ and\ \citenamefont
  {Hill}(1985)}]{parity_anomaly}%
  \BibitemOpen
  \bibfield  {author} {\bibinfo {author} {\bibfnamefont {S.}~\bibnamefont
  {Coleman}}\ and\ \bibinfo {author} {\bibfnamefont {B.}~\bibnamefont {Hill}},\
  }\href@noop {} {\bibfield  {journal} {\bibinfo  {journal} {Phys. Lett. B}\
  }\textbf {\bibinfo {volume} {159}},\ \bibinfo {pages} {184} (\bibinfo {year}
  {1985})}\BibitemShut {NoStop}%
\bibitem [{\citenamefont {Lee}(1986)}]{parity_anomaly_}%
  \BibitemOpen
  \bibfield  {author} {\bibinfo {author} {\bibfnamefont {T.}~\bibnamefont
  {Lee}},\ }\href@noop {} {\bibfield  {journal} {\bibinfo  {journal} {Phys.
  Lett. B}\ }\textbf {\bibinfo {volume} {171}},\ \bibinfo {pages} {247}
  (\bibinfo {year} {1986})}\BibitemShut {NoStop}%
\bibitem [{\citenamefont {Zubkov}(2018)}]{Zubkov2018a}%
  \BibitemOpen
  \bibfield  {author} {\bibinfo {author} {\bibfnamefont {M.~A.}\ \bibnamefont
  {Zubkov}},\ }\bibfield  {title} {\bibinfo {title} {{Momentum space topology
  of QCD}},\ }\href {https://doi.org/10.1016/j.aop.2018.04.016} {\bibfield
  {journal} {\bibinfo  {journal} {Annals Phys}\ }\textbf {\bibinfo {volume}
  {393}},\ \bibinfo {pages} {264} (\bibinfo {year} {2018})},\ \Eprint
  {https://arxiv.org/abs/1610.08041} {arXiv:1610.08041} \BibitemShut {NoStop}%
\bibitem [{\citenamefont {Zhang}\ and\ \citenamefont {Zubkov}(2019)}]{ZZ2019}%
  \BibitemOpen
  \bibfield  {author} {\bibinfo {author} {\bibfnamefont {C.~X.}\ \bibnamefont
  {Zhang}}\ and\ \bibinfo {author} {\bibfnamefont {M.~A.}\ \bibnamefont
  {Zubkov}},\ }\href@noop {} {\bibinfo {title} {{Influence of interactions on
  the anomalous quantum Hall effect}}} (\bibinfo {year} {2019}),\ \Eprint
  {https://arxiv.org/abs/arXiv:1902.06545} {arXiv:arXiv:1902.06545
  [cond-mat.mes-hall]} \BibitemShut {NoStop}%
\bibitem [{\citenamefont {Mera}(2017)}]{mera2017topological}%
  \BibitemOpen
  \bibfield  {author} {\bibinfo {author} {\bibfnamefont {B.}~\bibnamefont
  {Mera}},\ }\href@noop {} {\bibinfo {title} {Topological response of gapped
  fermions to a $\text{U}(1)$ gauge field}} (\bibinfo {year} {2017}),\ \Eprint
  {https://arxiv.org/abs/1705.04394} {arXiv:1705.04394 [cond-mat.str-el]}
  \BibitemShut {NoStop}%
\bibitem [{\citenamefont {Matsuyama}(1987)}]{Matsuyama1987a}%
  \BibitemOpen
  \bibfield  {author} {\bibinfo {author} {\bibfnamefont {T.}~\bibnamefont
  {Matsuyama}},\ }\bibfield  {title} {\bibinfo {title} {{Quantization of
  Conductivity Induced by Topological Structure of Energy Momentum Space in
  Generalized {QED} in Three-dimensions}},\ }\href@noop {} {\bibfield
  {journal} {\bibinfo  {journal} {Prog. Theor. Phys}\ }\textbf {\bibinfo
  {volume} {77}},\ \bibinfo {pages} {711} (\bibinfo {year} {1987})}\BibitemShut
  {NoStop}%
\bibitem [{\citenamefont {Zubkov}\ and\ \citenamefont {Wu}(2019)}]{ZW2019}%
  \BibitemOpen
  \bibfield  {author} {\bibinfo {author} {\bibfnamefont {M.~A.}\ \bibnamefont
  {Zubkov}}\ and\ \bibinfo {author} {\bibfnamefont {X.}~\bibnamefont {Wu}},\
  }\href@noop {} {\bibinfo {title} {{Topological invariant in terms of the
  Green functions for the Quantum Hall Effect in the presence of varying
  magnetic field}}} (\bibinfo {year} {2019}),\ \Eprint
  {https://arxiv.org/abs/arXiv:1901.06661} {arXiv:arXiv:1901.06661
  [cond-mat.mes-hall]} \BibitemShut {NoStop}%
\bibitem [{\citenamefont {Zhang}\ and\ \citenamefont {Zubkov}(2022)}]{ZZ2022}%
  \BibitemOpen
  \bibfield  {author} {\bibinfo {author} {\bibfnamefont {C.}~\bibnamefont
  {Zhang}}\ and\ \bibinfo {author} {\bibfnamefont {M.}~\bibnamefont {Zubkov}},\
  }\bibfield  {title} {\bibinfo {title} {Influence of interactions on integer
  quantum hall effect},\ }\href@noop {} {\bibfield  {journal} {\bibinfo
  {journal} {Annals of Physics}\ }\textbf {\bibinfo {volume} {444}},\ \bibinfo
  {pages} {169016} (\bibinfo {year} {2022})}\BibitemShut {NoStop}%
\bibitem [{\citenamefont {Khaidukov}\ and\ \citenamefont
  {Zubkov}(2017)}]{Khaidukov2017}%
  \BibitemOpen
  \bibfield  {author} {\bibinfo {author} {\bibfnamefont {Z.~V.}\ \bibnamefont
  {Khaidukov}}\ and\ \bibinfo {author} {\bibfnamefont {M.~A.}\ \bibnamefont
  {Zubkov}},\ }\bibfield  {title} {\bibinfo {title} {{Chiral Separation Effect
  in lattice regularization}},\ }\href
  {https://doi.org/10.1103/PhysRevD.95.074502} {\bibfield  {journal} {\bibinfo
  {journal} {Phys. Rev. D}\ }\textbf {\bibinfo {volume} {95}},\ \bibinfo
  {pages} {074502} (\bibinfo {year} {2017})},\ \Eprint
  {https://arxiv.org/abs/1701.03368} {arXiv:1701.03368} \BibitemShut {NoStop}%
\bibitem [{\citenamefont {Zubkov}\ and\ \citenamefont
  {Khaidukov}(2017)}]{Zubkov2017}%
  \BibitemOpen
  \bibfield  {author} {\bibinfo {author} {\bibfnamefont {M.~A.}\ \bibnamefont
  {Zubkov}}\ and\ \bibinfo {author} {\bibfnamefont {Z.~V.}\ \bibnamefont
  {Khaidukov}},\ }\bibfield  {title} {\bibinfo {title} {{Topology of the
  momentum space, Wigner transformations, and a chiral anomaly in lattice
  models}},\ }\href {https://doi.org/10.1134/S0021364017150139} {\bibfield
  {journal} {\bibinfo  {journal} {JETP Lett.}\ }\textbf {\bibinfo {volume}
  {106}},\ \bibinfo {pages} {166} (\bibinfo {year} {2017})},\ \bibinfo {note}
  {[Pisma Zh. Eksp. Teor. Fiz. {\bf 106} no.3, 166] (2017)}\BibitemShut
  {NoStop}%
\bibitem [{\citenamefont {Suleymanov}\ and\ \citenamefont
  {Zubkov}(2020)}]{SuleymanovZubkov2020}%
  \BibitemOpen
  \bibfield  {author} {\bibinfo {author} {\bibfnamefont {M.}~\bibnamefont
  {Suleymanov}}\ and\ \bibinfo {author} {\bibfnamefont {M.~A.}\ \bibnamefont
  {Zubkov}},\ }\bibfield  {title} {\bibinfo {title} {Chiral separation effect
  in nonhomogeneous systems},\ }\href
  {https://doi.org/10.1103/PhysRevD.102.076019} {\bibfield  {journal} {\bibinfo
   {journal} {Physical Review D}\ }\textbf {\bibinfo {volume} {102}},\ \bibinfo
  {pages} {076019} (\bibinfo {year} {2020})}\BibitemShut {NoStop}%
\bibitem [{\citenamefont {Zubkov}\ and\ \citenamefont
  {Abramchuk}(2023)}]{ZA2023}%
  \BibitemOpen
  \bibfield  {author} {\bibinfo {author} {\bibfnamefont {M.}~\bibnamefont
  {Zubkov}}\ and\ \bibinfo {author} {\bibfnamefont {R.~A.}\ \bibnamefont
  {Abramchuk}},\ }\bibfield  {title} {\bibinfo {title} {Effect of interactions
  on the topological expression for the chiral separation effect},\ }\href@noop
  {} {\bibfield  {journal} {\bibinfo  {journal} {Physical Review D}\ }\textbf
  {\bibinfo {volume} {107}},\ \bibinfo {pages} {094021} (\bibinfo {year}
  {2023})}\BibitemShut {NoStop}%
\bibitem [{\citenamefont {Zubkov}(2024)}]{Z2024}%
  \BibitemOpen
  \bibfield  {author} {\bibinfo {author} {\bibfnamefont {M.}~\bibnamefont
  {Zubkov}},\ }\bibfield  {title} {\bibinfo {title} {Weyl orbits as probe of
  chiral separation effect in magnetic weyl semimetals},\ }\href@noop {}
  {\bibfield  {journal} {\bibinfo  {journal} {Journal of Physics: Condensed
  Matter}\ }\textbf {\bibinfo {volume} {36}},\ \bibinfo {pages} {415501}
  (\bibinfo {year} {2024})}\BibitemShut {NoStop}%
\bibitem [{\citenamefont {Xavier}\ and\ \citenamefont {Zubkov}(2024)}]{XZ2024}%
  \BibitemOpen
  \bibfield  {author} {\bibinfo {author} {\bibfnamefont {P.~D.}\ \bibnamefont
  {Xavier}}\ and\ \bibinfo {author} {\bibfnamefont {M.}~\bibnamefont
  {Zubkov}},\ }\bibfield  {title} {\bibinfo {title} {Generalized wigner-weyl
  calculus for gauge theory and non-dissipative transport},\ }\href@noop {}
  {\bibfield  {journal} {\bibinfo  {journal} {arXiv preprint arXiv:2410.06952,
  to appear in Physical Review D}\ } (\bibinfo {year} {2024})}\BibitemShut
  {NoStop}%
\bibitem [{\citenamefont {Griffiths}\ and\ \citenamefont
  {Schroeter}(2018)}]{griffiths2018introduction}%
  \BibitemOpen
  \bibfield  {author} {\bibinfo {author} {\bibfnamefont {D.~J.}\ \bibnamefont
  {Griffiths}}\ and\ \bibinfo {author} {\bibfnamefont {D.~F.}\ \bibnamefont
  {Schroeter}},\ }\href@noop {} {\emph {\bibinfo {title} {Introduction to
  Quantum Mechanics}}},\ \bibinfo {edition} {3rd}\ ed.\ (\bibinfo  {publisher}
  {Cambridge University Press, Cambridge},\ \bibinfo {year} {2018})\BibitemShut
  {NoStop}%
\bibitem [{\citenamefont {Hillery}\ \emph {et~al.}(1984)\citenamefont
  {Hillery}, \citenamefont {O'Connell}, \citenamefont {Scully},\ and\
  \citenamefont {Wigner}}]{hillery1984distribution}%
  \BibitemOpen
  \bibfield  {author} {\bibinfo {author} {\bibfnamefont {M.}~\bibnamefont
  {Hillery}}, \bibinfo {author} {\bibfnamefont {R.~F.}\ \bibnamefont
  {O'Connell}}, \bibinfo {author} {\bibfnamefont {M.~O.}\ \bibnamefont
  {Scully}},\ and\ \bibinfo {author} {\bibfnamefont {E.~P.}\ \bibnamefont
  {Wigner}},\ }\bibfield  {title} {\bibinfo {title} {Distribution functions in
  physics: Fundamentals},\ }\href
  {https://doi.org/10.1016/0370-1573(84)90160-1} {\bibfield  {journal}
  {\bibinfo  {journal} {Physics Reports}\ }\textbf {\bibinfo {volume} {106}},\
  \bibinfo {pages} {121} (\bibinfo {year} {1984})}\BibitemShut {NoStop}%
\bibitem [{\citenamefont {Moyal}(1949)}]{moyal1949quantum}%
  \BibitemOpen
  \bibfield  {author} {\bibinfo {author} {\bibfnamefont {J.~E.}\ \bibnamefont
  {Moyal}},\ }\bibfield  {title} {\bibinfo {title} {Quantum mechanics as a
  statistical theory},\ }\href {https://doi.org/10.1017/S0305004100000487}
  {\bibfield  {journal} {\bibinfo  {journal} {Mathematical Proceedings of the
  Cambridge Philosophical Society}\ }\textbf {\bibinfo {volume} {45}},\
  \bibinfo {pages} {99} (\bibinfo {year} {1949})}\BibitemShut {NoStop}%
\bibitem [{\citenamefont {Zachos}\ \emph {et~al.}(2005)\citenamefont {Zachos},
  \citenamefont {Fairlie},\ and\ \citenamefont
  {Curtright}}]{zachos2005quantum}%
  \BibitemOpen
  \bibfield  {author} {\bibinfo {author} {\bibfnamefont {C.}~\bibnamefont
  {Zachos}}, \bibinfo {author} {\bibfnamefont {D.}~\bibnamefont {Fairlie}},\
  and\ \bibinfo {author} {\bibfnamefont {T.}~\bibnamefont {Curtright}},\ }\href
  {https://doi.org/10.1142/9789812799623} {\emph {\bibinfo {title} {Quantum
  Mechanics in Phase Space: An Overview with Selected Papers}}}\ (\bibinfo
  {publisher} {World Scientific, London},\ \bibinfo {year} {2005})\BibitemShut
  {NoStop}%
\bibitem [{\citenamefont {Zubkov}\ and\ \citenamefont
  {Wu}(2020)}]{zubkov2020topological}%
  \BibitemOpen
  \bibfield  {author} {\bibinfo {author} {\bibfnamefont {M.}~\bibnamefont
  {Zubkov}}\ and\ \bibinfo {author} {\bibfnamefont {X.}~\bibnamefont {Wu}},\
  }\bibfield  {title} {\bibinfo {title} {Topological invariant in terms of the
  green functions for the quantum hall effect in the presence of varying
  magnetic field},\ }\href@noop {} {\bibfield  {journal} {\bibinfo  {journal}
  {Annals of Physics}\ }\textbf {\bibinfo {volume} {418}},\ \bibinfo {pages}
  {168179} (\bibinfo {year} {2020})}\BibitemShut {NoStop}%
\end{thebibliography}%


\begin{thebibliography}{1}%
\makeatletter
\providecommand \@ifxundefined [1]{%
 \@ifx{#1\undefined}
}%
\providecommand \@ifnum [1]{%
 \ifnum #1\expandafter \@firstoftwo
 \else \expandafter \@secondoftwo
 \fi
}%
\providecommand \@ifx [1]{%
 \ifx #1\expandafter \@firstoftwo
 \else \expandafter \@secondoftwo
 \fi
}%
\providecommand \natexlab [1]{#1}%
\providecommand \enquote  [1]{``#1''}%
\providecommand \bibnamefont  [1]{#1}%
\providecommand \bibfnamefont [1]{#1}%
\providecommand \citenamefont [1]{#1}%
\providecommand \href@noop [0]{\@secondoftwo}%
\providecommand \href [0]{\begingroup \@sanitize@url \@href}%
\providecommand \@href[1]{\@@startlink{#1}\@@href}%
\providecommand \@@href[1]{\endgroup#1\@@endlink}%
\providecommand \@sanitize@url [0]{\catcode `\\12\catcode `\$12\catcode
  `\&12\catcode `\#12\catcode `\^12\catcode `\_12\catcode `\%12\relax}%
\providecommand \@@startlink[1]{}%
\providecommand \@@endlink[0]{}%
\providecommand \url  [0]{\begingroup\@sanitize@url \@url }%
\providecommand \@url [1]{\endgroup\@href {#1}{\urlprefix }}%
\providecommand \urlprefix  [0]{URL }%
\providecommand \Eprint [0]{\href }%
\providecommand \doibase [0]{https://doi.org/}%
\providecommand \selectlanguage [0]{\@gobble}%
\providecommand \bibinfo  [0]{\@secondoftwo}%
\providecommand \bibfield  [0]{\@secondoftwo}%
\providecommand \translation [1]{[#1]}%
\providecommand \BibitemOpen [0]{}%
\providecommand \bibitemStop [0]{}%
\providecommand \bibitemNoStop [0]{.\EOS\space}%
\providecommand \EOS [0]{\spacefactor3000\relax}%
\providecommand \BibitemShut  [1]{\csname bibitem#1\endcsname}%
\let\auto@bib@innerbib\@empty
\bibitem [{\citenamefont {Zubkov}\ and\ \citenamefont
  {Wu}(2020)}]{zubkov2020topological}%
  \BibitemOpen
  \bibfield  {author} {\bibinfo {author} {\bibfnamefont {M.}~\bibnamefont
  {Zubkov}}\ and\ \bibinfo {author} {\bibfnamefont {X.}~\bibnamefont {Wu}},\
  }\bibfield  {title} {\bibinfo {title} {Topological invariant in terms of the
  green functions for the quantum hall effect in the presence of varying
  magnetic field},\ }\href@noop {} {\bibfield  {journal} {\bibinfo  {journal}
  {Annals of Physics}\ }\textbf {\bibinfo {volume} {418}},\ \bibinfo {pages}
  {168179} (\bibinfo {year} {2020})}\BibitemShut {NoStop}%
\end{thebibliography}%
\end{document}